\documentclass[fleqn,usenatbib]{mnras}
\usepackage[utf8]{inputenc}
\usepackage{threeparttable}
\usepackage{amsmath}
\usepackage{graphicx}
\usepackage{epstopdf}
\usepackage{threeparttable}
\usepackage{graphicx, color}
\usepackage{physics}
\usepackage{bm}
\usepackage{soul}
\usepackage{amssymb}
\usepackage{upgreek}
\usepackage{hyperref}
\usepackage{ulem}

\usepackage{dcolumn}

\newcommand{\ay}{$\vb{A}_y$}

\newcommand{\tlc}{\tau_{\rm lc}}

\graphicspath{{figures/}} 

\title[Radiation from Plasmoids and Plasmoid Mergers]  
{Radiative Properties of Plasmoids and Plasmoid Mergers in Magnetic Reconnection}

\author[H. Zhang, L. Dong and D. Giannios]{
Haocheng Zhang$^{1,2}$, Lingyi Dong$^{3}$, Dimitrios Giannios$^{3}$\\
$^{1}$ University of Maryland Baltimore County, Baltimore, MD 21250, USA \\
$^{2}$ NASA Goddard Space Flight Center, Greenbelt, MD 20771, USA \\
$^{3}$ Department of Physics, Purdue University, West Lafayette, IN 47907, USA 
}

\begin{document}

\date{Received.../Accepted...}

\pagerange{\pageref{firstpage}--\pageref{lastpage}} \pubyear{2022}

\maketitle

\label{firstpage}

\begin{abstract}
Magnetic reconnection is often considered as the primary particle acceleration mechanism in a magnetized blazar zone environment. The majority of radiation in the reconnection layer comes from plasmoids and their mergers. In particular, plasmoid mergers can produce strong multi-wavelength flares and major variations in synchrotron polarization signatures. However, radiative properties of plasmoid mergers have not been well explored due to difficulties in tracking the merging processes. Here we use an image processing method that combines the magnetic vector potential and density to identify isolated and merging plasmoids. We find that this method can clearly distinguish radiation contributions from isolated plasmoids, merging plasmoids, and the primary current sheet of reconnection. This new method enables us to study the radiative properties of plasmoids and mergers statistically. Our results show that isolated plasmoids have similar emissivity regardless of their sizes, and they generally have nonzero polarization degree (PD) due to their quasi-circular shape. Flares due to plasmoid mergers have relative amplitudes that are anti-proportional to the size ratio of the plasmoids participating in the mergers. Finally, only mergers between plasmoids of comparable sizes (width ratio $\lesssim 5$) can lead to significant spectral hardening and polarization angle (PA) variations; the amplitude of the PA variations is between 0 and $180^{\circ}$ and has a mean value of $90^{\circ}$. Our analyses on 2D simulations can pave the way for future analyses and machine learning techniques on radiative properties of 3D magnetic reconnection simulations.
\end{abstract}


\section{Introduction}\label{sec:intro}

Blazars are relativistic collimated plasma outflows, also known as jets, that point very close to our line of sight. These jets are launched from accreting supermassive black holes with very high magnetic energy \citep{Blandford1977,Tchekhovskoy2010}. Around sub-parsec distance to a few parsecs, there is an unresolved region that emits highly variable nonthermal-dominated multi-wavelength emission, often referred to as the blazar zone. The variability time scale can reach as short as a few minutes \citep{Albert2007,Aharonian2007,Ackermann2016}, indicating extremely fast particle acceleration. Multi-wavelength campaigns often find that the spectral index of some observational bands hardens when those bands exhibit flares \citep{Giommi1990,Abdo2010,Krauss2016}, referred to as the ``harder-when-brighter'' trend. This trend further supports that these blazar flares are indeed driven by particle acceleration in situ in the blazar zone. The radio to optical blazar emission, in some cases up to X-ray bands, is dominated by the synchrotron emission from ultra-relativistic electrons, evident by the observed high PD \citep{Scarpa1997,Smith2017,Lister2018}. Observations find that the polarization signatures can also evolve in time \citep{Chandra2015,Covino2015,Itoh2016,Jorstad2022}. Very interestingly, optical polarization monitoring programs find that PA swings, defined as angle rotating continuously in one direction for $\gtrsim 90^{\circ}$, are typically accompanied by multi-wavelength flux variability \citep{Marscher2010,Larionov2013,Blinov2016,Blinov2018}, hinting the co-evolution of magnetic fields during particle acceleration \citep{ZHC2015,ZHC2018}. Such PA swings have been reported in soft X-rays by the recently launched IXPE as well \citep{DiGesu2023}. It is therefore believed that the blazar zone is probably magnetized, where the jet dissipates a significant amount of its magnetic energy to accelerate particles. Although the radio to optical emission is dominated by nonthermal electrons, the blazar jet may contain a proton population, which is evident by the detection of a very high energy neutrino event coinciding with the blazar TXS~0506+056 flare \citep{IceCube2018}. Understanding the physical conditions of the blazar zone and particle acceleration mechanisms therein is of critical importance to the time-domain and multi-messenger astronomy.

Magnetic reconnection is the primary plasma physical mechanism to accelerate nonthermal particles in a magnetized environment. This plasma physical process happens when two oppositely oriented magnetic field lines come close to each other, break and rejoin to rearrange the magnetic topology, and dissipate a large amount of the magnetic energy between the field lines. Such magnetic field topology can form during magnetic instabilities such as kink instabilities or in a striped jet configuration \citep{Begelman1998,Giannios2006,Giannios2019,Alves2018,Davelaar2020}. Previous particle-in-cell (PIC) simulations show that reconnection can efficiently accelerate both electrons and protons \citep{Sironi2014,Sironi2015,Guo2014,Guo2016,Guo2021,Werner2016,Werner2018,Li2018,Cerutti2012b,ZSG2021,Chernoglazov2023}, making it an ideal particle acceleration mechanism for hadronic blazar models. Particles are accelerated via Fermi-like mechanisms \citep{Guo2023,French2023}, or they may be accelerated directly in the electric field generated by reconnection \citep{Sironi2022}. The accelerated particle spectral indices generally depend on the magnetization factor, which is the ratio of magnetic energy over enthalpy, $\sigma$, and the guide field strength, which is the magnetic field component that is perpendicular to the anti-parallel field lines, $B_g$ \citep{Guo2015,Sironi2015}. Combined PIC and radiation transfer simulations find that reconnection typically leads to the harder-when-brighter trend during flares and polarization can strongly vary as well \citep{ZHC2020,ZHC2021b}. When the 2D reconnection starts, the current sheet, the thin layer between the anti-parallel magnetic field lines, can generate a series of quasi-circular structures with concentrated magnetic fields and nonthermal particles, called the plasmoids \citep{Sironi2014,Guo2014,Petropoulou2018,Hakobyan2021}. They contribute the majority of nonthermal emission \citep{ZHC2018,Christie2019}. Plasmoids move in the outflow direction of the reconnection region. Some plasmoids can reach relativistic speeds; if the outflow direction is along the line of sight, they can lead to extra relativistic Doppler boosting (the jet-in-jet model) and produce ultra-fast flares \citep{Giannios2009}. Plasmoids can also collide into each other and merge, during which they can dissipate a large amount of their magnetic energy and accelerate additional nonthermal particles, resulting in flares and PA swings \citep{ZHC2020,ZHC2022}. Although 3D simulations suggest that reconnection can generate strong turbulence that results in more complicated structures, 2D slices along the guide field direction still generally show plasmoid structures \citep{Guo2021,Huang2016}. Given the very high computational cost, it is impossible to simulate a realistic blazar zone using combined PIC and radiation transfer to study the radiation signatures of magnetic reconnection. It is therefore important to understand the radiative properties of plasmoids in more detail to interpret the radiation signatures on realistic physical scales.

\citet{Sironi2016} is a pioneering study on the physical properties of plasmoids. Using the local saddle points of the magnetic vector potential, the paper segments plasmoids and finds that plasmoids are generally self-similar structures. This paves the way for modeling radiation signatures of magnetic reconnection on the scale of realistic blazar zones. \citet{Christie2019} models the multi-wavelength blazar radiation from reconnection based on the above findings. However, the above segmentation method has a major drawback: by definition, adjacent plasmoids or groups of plasmoids share a common saddle point. Thus all plasmoids appear to be merging, making it impossible to distinguish isolated and merging plasmoids and study their respective radiative properties. Moreover, the radiation signatures in \citet{Christie2019} are not computed directly from PIC simulations, but based on simplified modeling assumptions inspired by \citet{Sironi2016}. For instance, one of the key assumptions is to consider plasmoids as unresolved concentrations of magnetic fields and particles, preventing the study of polarization signatures. To further study the radiative properties under first principles and enable direct comparison with 3D simulations and in future large-scale hybrid simulations, we need to better segment plasmoids and their mergers to explore their radiative properties statistically.

Here we introduce a new segmentation method using image processing techniques to identify isolated plasmoids and plasmoid mergers in combined 2D PIC and radiation transfer simulations of magnetic reconnection. The boundaries of plasmoids are determined by both the magnetic vector potential and particle density. Our approach can distinguish radiation contributions from plasmoids, mergers, and the primary current sheet. By statistically studying hundreds of plasmoids and tens of mergers from nine simulations with different physical parameters, we are able to examine their radiative properties, including polarization signatures. Section \ref{sec:method} presents our simulation setup and methodology, Section \ref{sec:overall} shows the overall radiation signatures based on our segmentation method, Sections \ref{sec:isolated} and \ref{sec:merger} describe radiative properties of isolated and merging plasmoids, respectively. We summarize and discuss our results in Section \ref{sec:summary}.


\section{Methodology}
\label{sec:method}

The goal of this paper is to study the radiative properties of isolated and merging plasmoids statistically. Thus we include radiative cooling effects for synchrotron and Compton scattering in our simulations \citep{ZHC2022}. \citet{Sironi2016} shows that the open boundary simulation is advantageous in generating a large number of plasmoids, but the number of plasmoid mergers, especially mergers between large plasmoids, is very limited. Here we choose periodic boundary conditions for our combined PIC and radiation transfer simulations, which guarantee that most of plasmoids will merge so as to boost the plasmoid merger statistics. We use image processing techniques to combine both the magnetic vector potential and plasma density to identify the boundary of plasmoids and their mergers. We find that our approach can clearly separate isolated plasmoids, plasmoids mergers, and the primary current sheet. Additionally, both isolated and merging plasmoid boundaries can nicely envelop their synchrotron radiation, ideal for studying the radiative properties. Our method can also track the evolution of both isolated and merging plasmoids. We are able to segment nearly one thousand plasmoids and a few hundred mergers from nine simulations with various physical conditions. We find that the PA swing driven by mergers is due to the merging plasmoids rotating around each other, resulting in the newly accelerated particles at the merging site streaming along the curved magnetic field lines surrounding the merging plasmoids as found in previous works \citep{ZHC2018,ZHC2021b}. This section presents our simulation setups and segmentation methods. We also describe how we define the radiative properties of isolated and merging plasmoids in the statistical studies.

\subsection{Simulation setup}

\begin{table*}
\centering
\caption{Parameters of nine PIC simulation runs and corresponding isolated plasmoids and binary plasmoid mergers identified. Run 2 and 3 are for different electron magnetization factor $\sigma_e$; Run 4 and 5 are for different initial upstream temperatures $T_e$; Run 6 and 7 are for different guide fields $B_g/B_0$; Run 8 and 9 are for different cooling factors $C_{10^4}$. For each run, we also listed the number of isolated plasmoids and binary plasmoids mergers found between $t=1.0~\tlc$ and $t=3.0~\tlc$.}
\label{tab:params}
\begin{tabular}[c]{cccccccc}
\hline
Run Number & $\sigma_e$ &  $T_e$ & $B_g/B_0$ & $C_{10^4}$ & Plasmoids & Binary Mergers\\
\hline
Run 1 & $4.0\times10^4$ & 100 & 0.2 & 200 & 138 & 8\\
Run 2 & \boldmath$8.0\times10^4$ & 100 &  0.2 & 200 & 59 & 7\\
Run 3 & \boldmath$2.0\times10^4$ & 100 &  0.2 & 200 & 170 & 18\\
Run 4 & $4.0\times10^4$ & \textbf{200} & 0.2 &  200 & 163 & 15\\
Run 5 & $4.0\times10^4$ & \textbf{50} & 0.2 &  200 & 106 & 7\\
Run 6  & $4.0\times10^4$ & 100 &  \textbf{0.1} &  200 & 142 & 5\\
Run 7 & $4.0\times10^4$ & 100 & \textbf{0.3} & 200&  91 & 12\\
Run 8 & $4.0\times10^4$ & 100 &  0.2 & \textbf{400} & 151 & 15\\
Run 9 & $4.0\times10^4$ & 100 &  0.2 & \textbf{100} & 154 & 11\\
\hline
\end{tabular}
\end{table*}

Our combined PIC and polarized radiation transfer simulations follow closely to \citet{ZHC2020}. Here we only summarize the basic simulation setup and describe several key physical parameters in Run 1. Other runs have different parameters from Run 1 that are highlighted in Table \ref{tab:params}. We assume a pre-existing current sheet moving in the jet direction $z$ with a bulk Lorentz factor $\Gamma=10$. The observer is viewing perpendicular to the jet propagation direction along $y$ axis in the comoving frame, so that in the observer's frame the viewing angle is $1/\Gamma$ from the jet axis and the Doppler factor $\delta\equiv \Gamma$. In this way, PIC simulations are performed in the comoving frame, while the radiation transfer includes the bulk relativistic beaming effects when it calculates the observable signatures. Given that the blazar can have a proton population, we assume a proton-electron plasma with a realistic mass ratio, $m_p/m_e=1836$. The initial electron magnetization factor is $\sigma_e=4\times 10^4$, which is related to the total magnetization factor by $\sigma_e\sim 1836\sigma$, and we consider a guide field $B_g=0.2B_0$, where $B_0$ is the anti-parallel component of the magnetic field. Particles initially follow a Maxwell–J\"uttner distribution with the same upstream temperature for electrons and protons, $T_e=T_p=100m_ec^2/k_B$, where $k_B$ is the Boltzmann constant. Our analyses focus on the optical synchrotron emission, thus the upstream electrons cannot affect our results as their synchrotron critical energy is well below the optical band. We add a radiative reaction force to mimic the cooling effect, parameterized by $C_{10^4}$. This parameter describes the strength of the radiative cooling at $\gamma_e=10^4$, and a smaller value means stronger cooling. Our simulation grid size is $4096\times 2048$ in the $x$-$z$ plane with a physical size of $2L\times L$, where $L=8000 d_{e0}$ and $d_{e0}$ is the non-relativistic electron inertial length. This corresponds to $L\sim 656 d_e$, where $d_e=\sqrt{1+3T_e/(2m_ec^2)}d_{e0}\sim 12.2 d_{e0}$ is the upstream electron inertial length. Each cell has 100 particles. The simulation is performed with the \texttt{VPIC} code developed by \citet{Bowers2008}.

The magnetic fields and nonthermal particle spectra at each snapshot of the above simulation are post-processed with polarized radiation transfer code \texttt{3DPol} \citep{ZHC2014}. The particle spectra are obtained by counting the number of electrons with $\gamma-1$ in a certain bin. We use 100 bins between $10^{-4}\leq \gamma-1 \leq 10^6$, adequate to obtain smooth synchrotron spectrum. To obtain sufficient particle statistics for the spectrum, we aggregate $16\times 16$ PIC cells into one radiation transfer cell. Our studies thus do not consider plasmoids or their mergers smaller than $16\times 16$ PIC cells; radiation contribution from such small plasmoids is counted towards the primary current sheet. Since our simulation is 2D and we view the simulation domain in the $y$ direction, every PIC snapshot is post-processed to one snapshot of the polarized emission map. Since the radiation signatures in the observer's frame are different from the comoving frame in the Doppler-boosted flux only (PD and PA are unaffected), in the following analysis we show the comoving frame radiation signatures. We normalize the initial anti-parallel magnetic field $B_0=0.1~\rm{G}$. All the radiation signatures shown in the following are in the optical band, which is approximately at the cooling break of the synchrotron spectrum. To closely track the variability, the magnetic field and particle data are dumped every $t=0.0125\tau_{lc}$ from \texttt{VPIC} to \texttt{3DPol}, where $\tau_{lc}$ is the light crossing time of the simulation domain in the $x$ direction.

\subsection{Segmentation Method}

\begin{figure}
\centering
\includegraphics[width=0.45\textwidth]{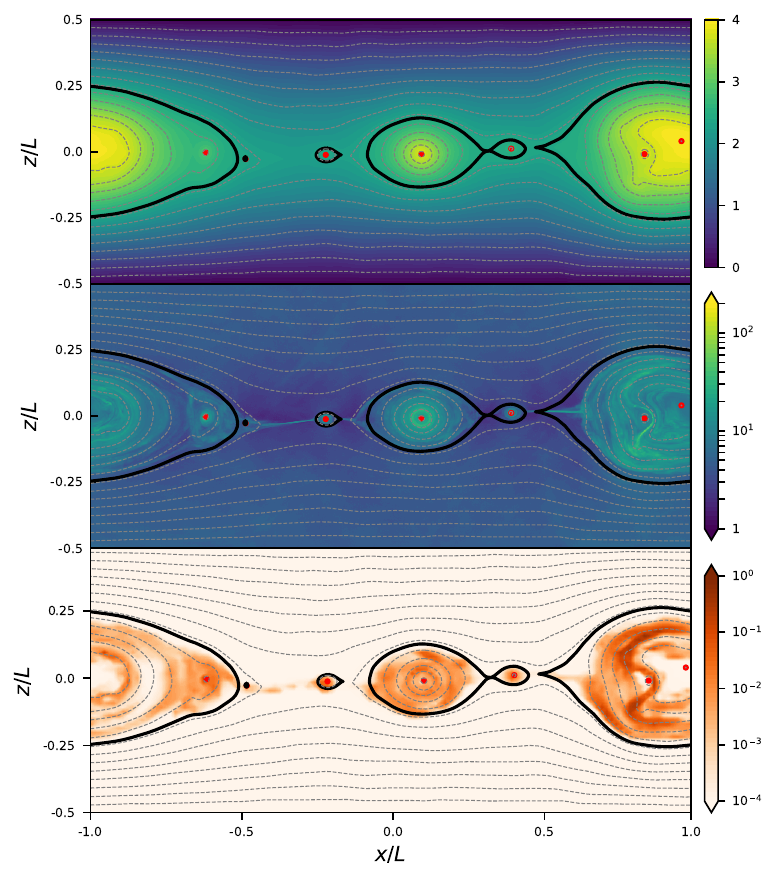}
\caption{One snapshot of the simulation showing the plasmoid segmentation result. Three panels from top to bottom are the \ay \ distribution, particle density distribution and optical emission map in log scale. The current sheet is along the $x$-axis, the line of sight in the comoving frame is along the $y$-axis, and the bulk motion of the simulation domain is along the $z$-axis. The thin gray dashed curves represent magnetic field lines in the simulation. Red dots represent O-points, and black contour lines show the isolated plasmoids and plasmoid mergers from our segmentation method. It is clear that there is an ongoing merger involving three plasmoids at the periodic boundary of the simulation domain (represented by the three O-points within the merger) and three isolated plasmoids near the center. Two of those three isolated plasmoids (between $x/L=0$ and $x/L=0.5$) are about to merge.}
\label{fig:method}
\end{figure}

We use the magnetic vector potential $\vb{A}$ to determine the location of plasmoids as in previous works \citep{Sironi2016,Banesh2020,Hakobyan2021}. Since our simulations are 2D in the $x$-$z$ plane, it is simply $A_y$. Local maxima and minima in $A_y$, often referred to as O-points, define the topological centers of plasmoids. For every O-point identified in one snapshot, there exists a specific $A_y$ isovalue whose contour is maximized but encloses only this single O-point. By implementing a binary search algorithm, we can find the $A_y$ isovalues for every detected O-point in every snapshot of our simulations. We note that this value is equivalent to the magnetic vector potential value at the adjacent saddle points (i.e., X-points) of each plasmoid as in the previous works \citep{Sironi2016,Banesh2020,Hakobyan2021}.

Plasmoids selected by $A_y$ alone are connected to each other (or other groups of plasmoids) via their adjacent X-points. However, they are not necessarily in the merging process. We choose the particle density as an additional parameter for the plasmoid boundary. However, the particle density is not exactly the same for all plasmoids, nor is it uniform inside a specific plasmoid. \citet{Sironi2016} shows that the average particle density in plasmoids is similar regardless of plasmoid sizes, and it is considerably higher than the inflow density. Inspired by this, we use an imaging processing method called the Otsu's threshold to distinguish plasmoids that have similarly high particle density and the rest of the simulation domain. This method can separate pixels of the grayscale plot of the particle density into a foreground class (plasmoids) and a background class (rest of the simulation domain) by a threshold particle density value, such that this threshold can minimize the intra-class variance of particle density in the pixels \citep{Otsu1979}. In this way, the simulation domain is segmented into many patches of foreground pixels that have irregular shapes. We then check if two or more adjacent O-points share a common foreground patch: if so, they are an ongoing merger. In this situation, we ignore their individual $A_y$ contours, but find the $A_y$ isovalue whose contour is maximized but encloses only all O-points participating the merger. We remind the readers that although our segmentation method can identify plasmoids and mergers of arbitrary sizes, our studies of physical and radiative properties only consider plasmoids and mergers larger than $16\times 16$ PIC cells.

\subsection{Tracking Method for Isolated and Merging Plasmoids}

\begin{figure*}
\centering
\includegraphics[width=0.95\textwidth]{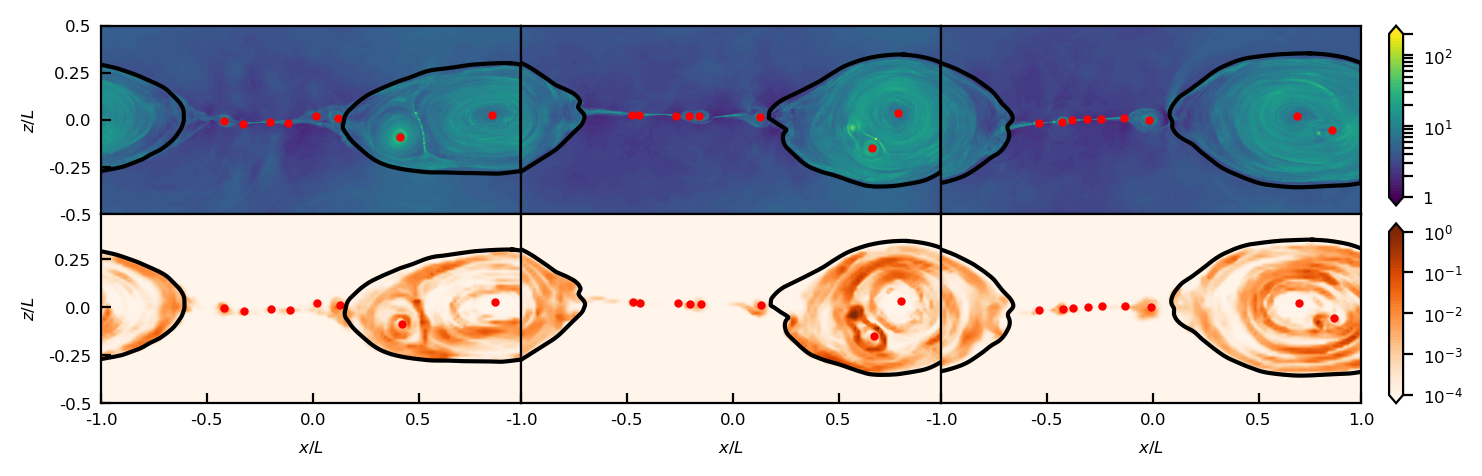}
\caption{Snapshots of a plasmoids merger in Run 1, whose boundary is marked by the black contour. The three snapshots are selected a few time steps after the beginning of the merger ($t=1.88\tau_{lc}$), near the peak of the flare ($t=2.22\tau_{lc}$), and right before the merger completes ($t=2.48\tau_{lc}$). Top and bottom panels are proton density and optical flux, respectively.}
\label{fig:snapshots_run1_idx45}
\end{figure*}

\begin{figure}
\centering
\includegraphics[width=0.45\textwidth]{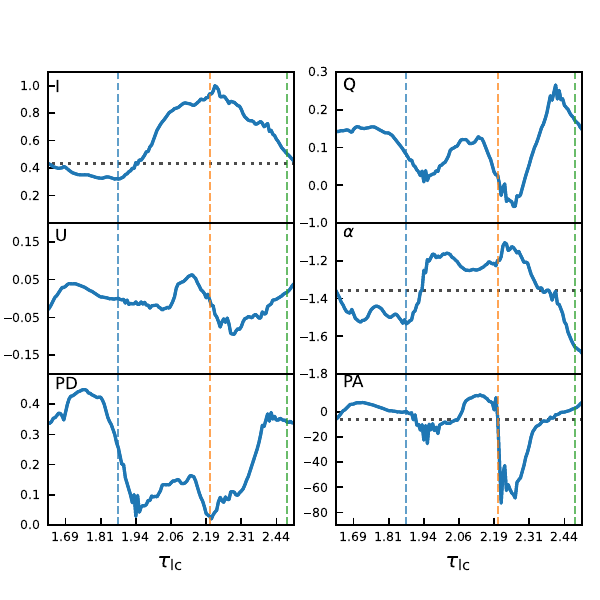}
\caption{Temporal evolution of radiation signatures of the merger shown in Figure~\ref{fig:snapshots_run1_idx45}. From left to right then top to bottom are the optical Stokes $I$, $Q$, $U$, optical spectral index $\alpha$, optical PD and PA. Vertical red dashed lines correspond to three selected snapshots. Gray dashed lines correspond to the values at the beginning of the merger. All curves are for the merger only (i.e., radiation and polarization within the black contour in Figure~\ref{fig:snapshots_run1_idx45}).}
\label{fig:curves_run1_idx45}
\end{figure}

\begin{figure}
\centering
\includegraphics[width=0.45\textwidth]{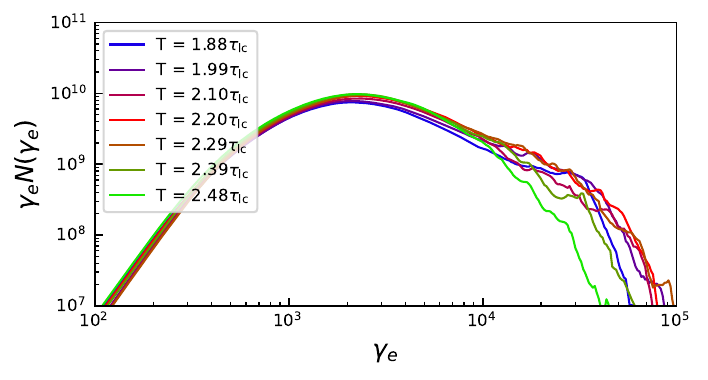}
\includegraphics[width=0.45\textwidth]{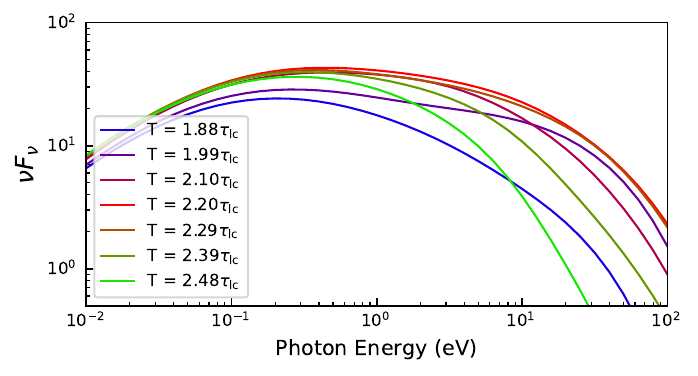}
\caption{Snapshots of particle and radiation spectra of the merging plasmoids within the black contour in Figure \ref{fig:snapshots_run1_idx45}. The upper panel plots the particle spectra and the bottom panel plots the radiation spectra in code units. Different colors represent seven snapshots that are evenly distributed in time.}
\label{fig:spectra_run1_idx45}
\end{figure}

The above segmentation can be done for every snapshot in all our simulations. Since we are interested in the variability, we need to track the motion and evolution of both isolated and merging plasmoids. For isolated plasmoids, we track the motion of their O-points. Since the speed of a plasmoid in our simulations can reach at most $\sim\sqrt{\sigma/(\sigma+1)} c$, corresponding to a maximal displacement of its O-point at $\sim 0.025 L$ in one time step. Therefore, if an O-point in two consecutive snapshots does not move more than the above value in the outflow direction, we label it as the same plasmoid. It is possible that new plasmoids can form between two snapshots. However, our snapshots are very dense in time, newly formed plasmoids are at most of a size of a couple of PIC cells. Thus our tracking method for isolated plasmoids larger than $16\times 16$ PIC cells will not be contaminated by newly formed plasmoids or secondary plasmoids generated at plasmoid merging sites. We then define the plasmoid width $w$ as its maximal size along the $z$-axis, and the plasmoid width along the primary current sheet as the maximal size along the $x$-axis. Other physical quantities such as the particle density $n$ and magnetic energy density $\varepsilon_B$ are the average values for cells within the plasmoid boundary. The Stokes $I$ (i.e., the luminosity) of an isolated plasmoid is the total luminosity from all radiation transfer cells enclosed by the plasmoid boundary, and the PD is calculated by $\sqrt{Q^2+U^2}/I$, where the Stokes $Q$ and $U$ are also summed over all cells in the plasmoid. We note that plasmoids can grow in time or merge with others and become newly isolated plasmoids after the merger. To account for such temporal evolution, an isolated plasmoid is counted as a new entry in the statistical study every ten snapshots if it stays isolated for more than ten snapshots; if two or more plasmoids complete a merging process, the post-merger isolated plasmoid is counted as a new entry immediately as well. Since we analyze the optical synchrotron emission in our statistical study, and $\sigma$ as well as cooling can affect whether the optical band is at the synchrotron spectral cooling break \citep{Sironi2016,ZHC2020}, we only consider Runs 1, 4, 5, 6, 7 that share the same $\sigma$ and cooling factor $C_{10^4}$ for isolated plasmoids.

Merging plasmoids are more complicated to track, because an ongoing merger often has an irregular shape that changes over time. In addition, mergers can also move in space just like isolated plasmoids. Moreover, ongoing mergers may collide and merge with other isolated and/or merging plasmoids. The beginning of a merger is straightforward to determine: if we find two or more isolated plasmoids in a given snapshot are labeled as a merger in the next snapshot, then the latter snapshot is considered as the beginning of the merger. Although mergers can move in space, their speed is non-relativistic in our simulations: they cannot move more than a few PIC cells in one snapshot. Since we only track plasmoids that are at least $16\times 16$ PIC cells, if the center of a merger between two snapshots does not change more than 16 PIC cells, it is considered as the same merger. We continue to track the merger as long as there are more than one O-point in the merger. Thus whether the merger collides with other isolated (or merging) plasmoids or not, it is considered as the same merger. As a result, it is possible that two ongoing mergers become one merger. However, mergers involving more than two plasmoids, either from the beginning or additional plasmoids join the ongoing merger, are ambiguous to determine the beginning state of the radiative properties. Therefore, even though our method can correctly track such complicated mergers, we only include binary mergers that involve only two plasmoids from the beginning to the end for the statistical study of variability. As listed in Table \ref{tab:params}, the total number of binary mergers is $\sim 100$, about 20\% of the total mergers that we track. Intuitively the merger is considered as complete when only one O-point remains in the merging region. This works well if the isolated plasmoid at the end of the merger is well resolved. But if the merging plasmoids are not well resolved, their O-points may disappear in one snapshot and reappear in a later snapshot. In this case, we continue to track a few more snapshots till the flare due to the merger ends. As we will see in the following, this does not affect how we evaluate the variability due to mergers, since that is determined by the maximal changes from the beginning of the merger, which is well determined. But it removes contamination from ``fake'' isolated plasmoids that are in fact mergers about to finish in the study of radiative properties of isolated plasmoids.

Figure \ref{fig:snapshots_run1_idx45} shows an example of a merger between two large plasmoids ($w\sim 0.5L$ and $w\sim 0.25L$). We can see that both the density and the optical flux are enhanced at the contact region of the two plasmoids in the first snapshot that is a few time steps after the beginning of the merger, indicating that the secondary current sheet has formed between the two plasmoids. The size of the smaller plasmoid shrinks and the shape of the secondary current sheet becomes irregular in the second snapshot that is near the peak of the flare. It is clear that the local flux in the secondary current sheet increases significantly. Interestingly, the O-point of the smaller plasmoid appears to rotate counterclockwise in the three snapshots. Previous works suggest that the PA swing is driven by newly accelerated particles in the secondary current sheet moving around the merging plasmoid. Our analysis now shows the physical reason for such movements: the merging plasmoids rotate around each other. However, this merger does not make a $\sim 180^{\circ}$ PA swing (Figure \ref{fig:curves_run1_idx45}). We can see that the Stokes $Q$ and $U$ reach nearly zero at the peak of the Stokes $I$, thus the PD becomes zero and PA is arbitrary. Right after the peak the PA flips to a different position, making an apparent $\sim 90^{\circ}$ rotation. However, this is not considered as a typical PA swing where the Stokes $Q$ and $U$ should follow approximately two sine functions that are off-phase by $\sim 90^{\circ}$ (equivalently, making a quasi-circle in the Stokes $Q$-$U$ plane around 0). Such behaviors originate from the very disordered magnetic field evolution during the merger. We will show an example of a typical PA swing in the next section. The optical photon index hardens during the flare as well \ref{fig:spectra_run1_idx45}. In Section \ref{sec:merger} we statistically study the flare amplitude, spectral hardening, and amplitude of the PA variation; the flare amplitude is defined by the ratio of the peak of the Stokes $I$ over its beginning value, while the other two are defined as the maximal change from the beginning value. Take this merger as an example, the flare amplitude is $\sim 2.5$, the spectral hardening is $\sim 0.3$, and the PA variation amplitude is $\sim -70^{\circ}$.

\section{Overall Segmentation Results}
\label{sec:overall}

\begin{figure}
\centering
\includegraphics[width=0.45\textwidth]{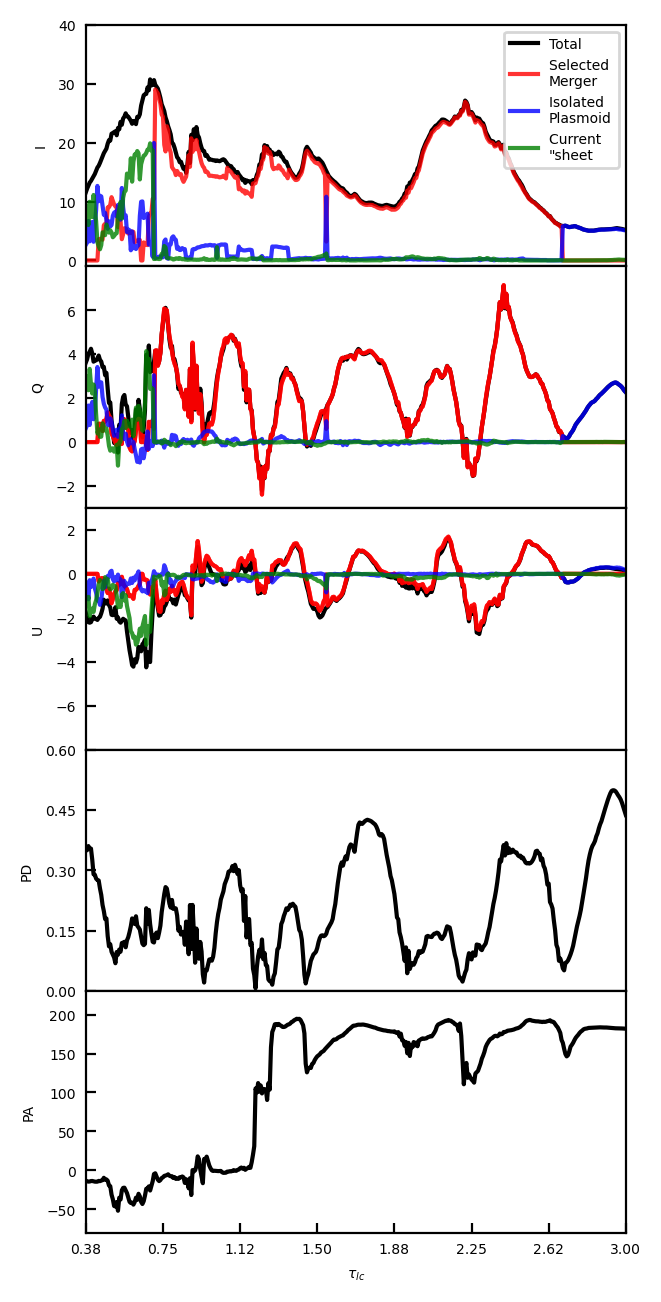}
\caption{From top to bottom panels are temporal evolution of optical Stokes $I$, $Q$, $U$, as well as PD, PA for Run 1, from $0.38\tau_{lc}$ to $3\tau_{lc}$. The black curve represents the overall evolution, while the red, blue, and green curves in the first three panels represent the contributions from all plasmoid mergers (including both binary mergers in Table~\ref{tab:params} and mergers involving multiple plasmoids), isolated plasmoids, and the primary current sheet, respectively. The contribution from the current sheet is obtained by subtracting the mergers and isolated plasmoid contributions from the total Stokes parameters. Stokes $I$, $Q$, $U$ are in code units, PD ranges from 0 to 1 (i.e., 0\% to 100\%), and PA is in degrees.}
\label{fig:run1}
\end{figure}

\begin{figure}
\centering
\includegraphics[width=0.45\textwidth]{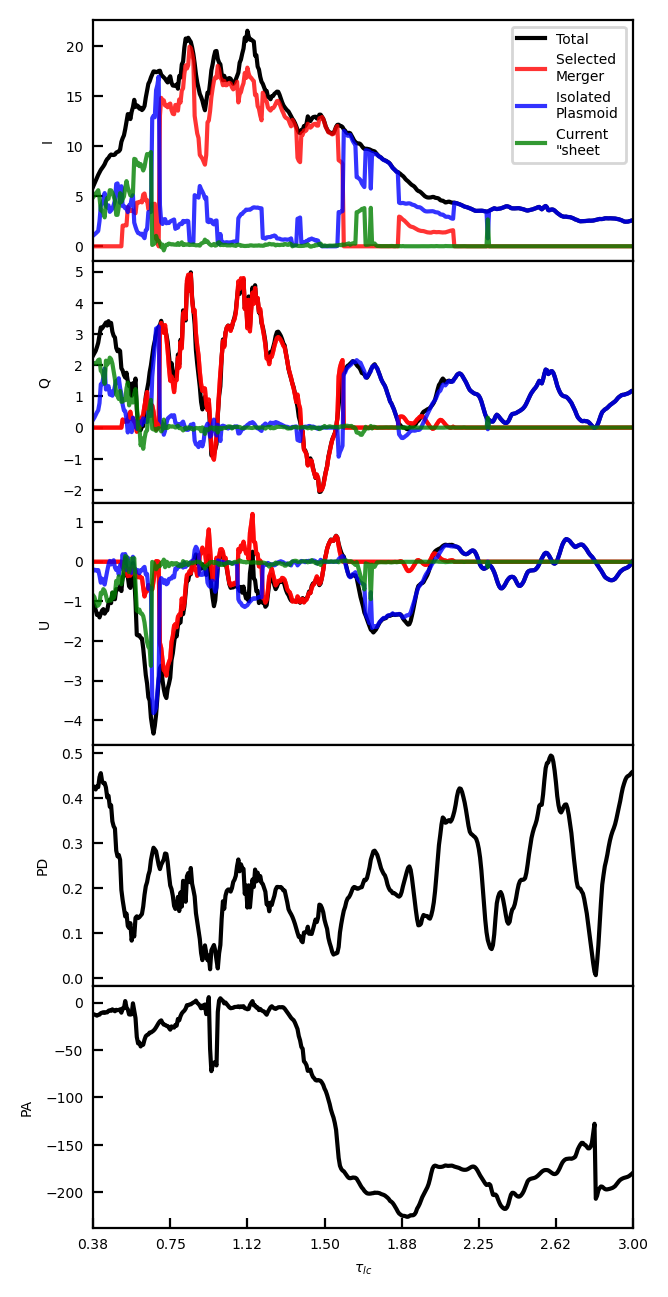}
\caption{Same as Figure \ref{fig:run1} but for Run 5}
\label{fig:run5}
\end{figure}

This section shows the overall radiation signatures from isolated and merging plasmoids as well as the primary current sheet. We find that the emission contribution from the primary current sheet is trivial compared to the isolated and merging plasmoids. Stokes $I$, $Q$, $U$ during flares are dominated by merging plasmoids, although this is probably biased due to the periodic boundary conditions, which can lead to a lot of mergers. All continuous PA rotations $\gtrsim 90^{\circ}$ are driven by mergers, and for swings $\gtrsim 180^{\circ}$, the Stokes $Q$ and $U$ follow roughly sine functions with $\sim 90^{\circ}$ difference in phase. Isolated plasmoids can have varying PD and PA due to the non-uniform distribution of nonthermal particles inside, but they cannot create $\gtrsim 90^{\circ}$ rotations.

Figure \ref{fig:run1} shows the overall temporal evolution of the optical Stokes $I$, $Q$, $U$ as well as PD and PA. The general patterns are similar to \citet{ZHC2020}: since the reconnection starts, the optical light curve (Stokes $I$) shows multiple flares till the reconnection gradually saturates at $t\gtrsim 2.5\tau_{lc}$. The first four relatively short flares before $t\sim 1.5\tau_{lc}$ are mostly due to isolated plasmoids and mergers in the reconnection layer, while the last giant flare between $t\sim 1.75\tau_{lc}$ and $t\sim 2.75\tau_{lc}$ results from a large binary merger. The first PA swing near $t=1.2 \tau_{lc}$ is due to mergers between multiple plasmoids, while the sudden PA drop around $t=2.25\tau_{lc}$ is due to the plasmoid merger shown in the previous section. Previous works have shown significant changes in PD and PA during flares \citep{ZHC2018,ZHC2020,ZHC2021b}, but here we also present how the Stokes $Q$ and $U$ vary to further explore the physical origin of polarization variations. As we can see, major rises and falls of the PD are highly correlated to flares of Stokes $Q$. The Stokes $U$, on the other hand, do not show a clear correlation with the PD or PA. The reason lies in the radiative properties of isolated plasmoids and mergers. Isolated plasmoids, due to their geometric shape, have a net polarization that is perpendicular to the reconnecting magnetic field lines (described in Section \ref{sec:isolated}). On the other hand, the secondary reconnection at the contact region of the merging plasmoids, generally leads to polarization along the reconnecting field lines. These two polarization directions give the positive and negative Stokes $Q$, while the Stokes $U$ results from particles accelerated at the secondary reconnection region that stream along the post-merger plasmoid, which have been partially cooled via radiation.

Our segmentation method allows us to distinguish different emission contributions. Figure \ref{fig:run1} shows that in Stokes $I$, $Q$, and $U$, plasmoid mergers contribute the most to emission, isolated plasmoids contribute a small portion, while the current sheet has a minimal contribution. We note that the dominance of plasmoid mergers in all Stokes parameters in our simulation is due to the periodic boundary conditions, which result in many more mergers compared to open boundaries as in \citet{Sironi2016}. Nonetheless, at an earlier stage ($0.5\tau_{lc}<t<1.5\tau_{lc}$), when both isolated and merging plasmoids exist in the simulation domain, merging plasmoids still contribute much more than isolated ones in radiation. This is because plasmoid mergers not only consume a significant amount of the magnetic energy in participating plasmoids, they also harden the particle spectra (described in Section \ref{sec:merger}), considerably boosting the optical flux. Radiation from the current sheet appears to be significant at the beginning of the reconnection ($t\lesssim 0.75 \tau_{lc}$); however, this is because we do not include plasmoids smaller than $16\times 16$ PIC cells, whose radiation is counted as from the current sheet. If we consider any plasmoids that are resolved by at least 2 PIC cells, then even the emission in the beginning of the simulation is dominated by isolated plasmoids (about 70\% to 80\%) and mergers (the rest), while the contribution from the current sheet is trivial. The Stokes $Q$ and $U$ are dominated by the mergers as well, further supporting that the polarization variations result from plasmoid mergers. We note that although mergers almost continuously happen during our simulations, there are a few snapshots where very few mergers are ongoing. In those snapshots, radiation from isolated plasmoids becomes significant, which leads to sudden changes in their contribution to the Stokes $I$, $Q$, $U$ as shown in Figure \ref{fig:run1}. Such sudden changes are simply due to the periodic boundaries that make mergers so frequent: if the simulation has open boundaries, where plasmoid mergers are much less frequent, the transition will appear smooth.

Figure \ref{fig:run5} presents a different run, in which a smooth large-amplitude PA swing happens. The general patterns are similar to Run 1, but this run does not have a major plasmoid merger near the end of the simulation. Instead, all plasmoid mergers are complete before $t\sim 1.75\tau_{lc}$, leaving two giant plasmoids that are far apart from each other (however, if the simulation continues for a longer time, they will merge eventually). As a result, the emission is dominated by isolated plasmoids later in the simulation. The distributions of nonthermal particles and magnetic fields can evolve in both plasmoids. They produce fluctuations in Stokes $Q$ and $U$, which lead to changes in the PA. But such fluctuations are not significant to make any PA swings $\gtrsim 90^{\circ}$. Since the two isolated plasmoids are the only dominating emission patches, the PD appears very high when Stokes $Q$ and $U$ rise. However, the reconnection has mostly stopped at this stage and the flux is very low, such high PDs are unlikely observable. The main difference of this run from Run 1 is the smooth large-amplitude PA swing around $t\sim 1.5\tau_{lc}$, which is due to a major plasmoid merger. We can see that the Stokes $Q$ follows roughly a cosine curve while the Stokes $U$ follows roughly a sine curve. The $Q$ and $U$ thus complete a quasi-circle around zero in the Stokes $Q$-$U$ plane (not shown here), a clear signature for PA swing found in observations. We will quantify the radiative properties of plasmoid mergers in Section \ref{sec:merger}. As we will see, the PA rotation amplitudes for plasmoid mergers can be stochastic, but they follow a general trend that depends on the size ratio of the merging plasmoids.

\section{Isolated Plasmoids}
\label{sec:isolated}

\begin{figure*}
\centering
\includegraphics[width=0.95\textwidth]{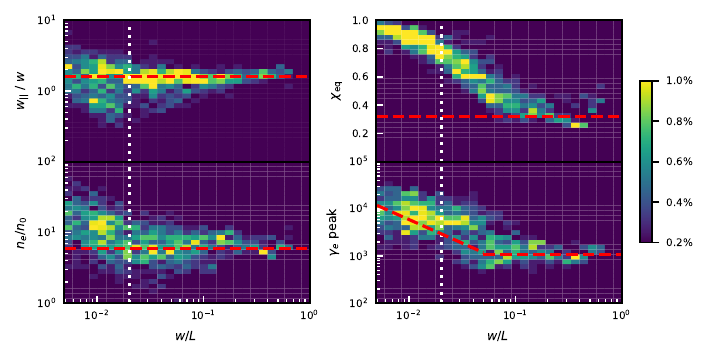}
\caption{2D histogram of the physical properties of isolated plasmoids vs. plasmoid width $w$. The red dashed line represents the theoretical values or trends discussed in this section. The white dotted line marks $w=0.02L$. Upper left panel: the ratio of the plasmoid widths parallel and perpendicular to the primary current sheet $w_{\parallel}/w$. Lower left panel: the ratio of the averaged electron density $n_e$ within the plasmoid over that in the upstream $n_0$. Upper right panel: the equipartition factor $\chi_{\rm eq}$ defined in this section. Lower right panel: the peak of the electron spectrum within the plasmoid.}
\label{fig:isolatedphys}
\end{figure*}

Here we present the physical and radiative properties of isolated plasmoids. We find that the majority of physical properties of isolated plasmoids are similar to those found in \citet{Sironi2016}. The difference is that our simulations include radiative cooling, thus the kinetic energy of particles in plasmoids can decrease, resulting in a lower equipartition factor towards the end of simulations (refer to Equation \ref{eqn:equipartition}). Similarly, the spectral peak of the nonthermal electrons decreases due to cooling (refer to \ref{fig:isolatedphys}). We find that all isolated plasmoids, regardless of their sizes, share approximately the same emissivity. They also exhibit a nonzero PD due to their quasi-circular geometric shape. We note that due to the periodic boundaries in our simulations, the number of isolated plasmoids is much less than that in \citet{Sironi2016}. Thus we use Runs 1, 4, 5, 6, and 7 that share the same magnetization factor $\sigma$ and cooling factor $C_{10^4}$ to increase the statistics. Unfortunately, our image processing techniques are still too computationally expensive to explore dependence on additional parameters. Nonetheless, key physical and radiative properties of plasmoids do not show clear dependence on their sizes, indicating that isolated plasmoids are indeed self-similar structures.

Figure \ref{fig:isolatedphys} shows the 2D histogram of four physical properties (in the comoving frame of the simulation domain) of isolated plasmoids against the plasmoid widths $w$ in the $z$ direction (i.e., perpendicular to the primary current sheet). As shown in the upper left panel, the plasmoids generally appear quasi-circular, where the width $w_{\parallel}$ along the primary current sheet (i.e., the $x$ axis) slightly larger than the width $w$ perpendicular to the primary current sheet, in agreement with \citet{Sironi2016}. The stretching along the primary current sheet is quite obvious in Figure \ref{fig:method} as well. The ratio $w_{\parallel}/w\sim 1.6$ (red dashed line in the upper left panel) is more pronounced for larger plasmoids that are well resolved. Small plasmoids of $w\lesssim 0.02L$ ($w=0.02L$ is marked by the white vertical dotted line in Figure \ref{fig:isolatedphys}) have a radius that contains only $\sim 20$ PIC cells. In this case, the systematic errors in the plasmoid contours can become non-trivial, thus there are more plasmoids $w_{\parallel}/w$ deviating from the 1.6 ratio.

The lower left panel of Figure \ref{fig:isolatedphys} shows the ratio of the average electron density in the plasmoids $n_e$ over that in the plasma inflow $n_0$. Our result is slightly higher than that in \citet{Sironi2016}: our total magnetization factor is $\sigma\sim 22$, but we find $n_e/n_0\sim 5$ (red dashed line in the lower left panel), which is the value for the $\sigma=50$ simulation in \citet{Sironi2016}. This is likely due to our different setups: we consider a proton-electron plasma with guide fields and radiative cooling, thus it is not an apples-to-apples comparison with \citet{Sironi2016}. Our value seems to match well with the $n_e/n_0\propto \sqrt{\sigma}$ scaling in \citet{Lyubarsky2005}; however, we are unable to verify the trend with more runs with different $\sigma$ due to the very computationally expensive image processing method. We note that our results are still consistent with the $n_e/n_0\gtrsim \Gamma$ argument in \citet{Sironi2016}, but we remind the readers that due to the periodic boundary conditions, our plasmoids are at most mildly relativistic, $\Gamma\lesssim 2$.

Other physical quantities (not shown) are in good agreement with \citet{Sironi2016}, just like $w_{\parallel}/w$ and $n_e/n_0$. However, our simulations include radiative cooling, hence the average electron kinetic energy in plasmoids can vary in time. This is best represented by the equipartition factor $\chi_{eq}$ shown in the upper right panel. We define $\chi_{eq}$ in a similar way as in \citet{Sironi2016},
\begin{equation}
\chi_{\rm eq}=\frac{\int \frac{\varepsilon_{\rm kin}}{1 + \varepsilon_{\rm B}/\varepsilon_{\rm kin}} dV}{\int \varepsilon_{\rm kin} dV}~~,
\label{eqn:equipartition}
\end{equation}
but we include both the kinetic energy density of protons and electrons in $\varepsilon_{\rm kin}$ since we use proton-electron plasma, while $\varepsilon_{\rm B}$ denotes the magnetic energy density. As stated in \citet{Sironi2016}, the strong kinetic dominance for small plasmoids can be attributed to that field lines generally wrap around large plasmoids, so that the small plasmoids preferentially lie in regions where the field lines are less dense, resulting in weaker magnetic fields. However, large plasmoids in our simulations reach a state that the magnetic energy is slightly dominating ($\chi_{eq}\sim 0.4$), different from $\chi_{eq}\sim 0.6$ in \citet{Sironi2016}. This is mainly due to radiative cooling. Previous works have found that in proton-electron plasma, the energy in nonthermal electrons and protons are comparable, i.e., $\varepsilon_{kin,e}\sim \varepsilon_{kin,p}$, and plasmoids will reach equipartition ($\varepsilon_{kin,e}+\varepsilon_{kin,p}\sim\varepsilon_{B}$) if there is no cooling. Since large plasmoids take a long time to grow, nonthermal electrons therein can radiatively cool and lose kinetic energy. Protons, due to their much longer cooling time, experience trivial cooling effects. As a simple estimate, if the electrons lose all their kinetic energy due to cooling, then the equipartition factor will be approximately $\chi_{eq}\sim \varepsilon_{kin,p}/(\varepsilon_{kin,p}+\varepsilon_B)\sim 1/3$. In practise, electrons will only lose part of their kinetic energy, thus the $\chi_{eq}$ for large plasmoids is higher, at roughly 0.4 (red dashed line) in Figure \ref{fig:isolatedphys} upper right panel.

The electron cooling is more clearly illustrated in the lower right panel of Figure \ref{fig:isolatedphys}, where we plot the spectral peak of the electron spectrum (peak refers to the peak in a log-log plot of $\gamma_e^2N_e(\gamma_e)$ against $\gamma_e$). We find that small plasmoids can reach $\gamma_{peak}\sim 10^4$, which is very close to the electron magnetization factor $\sigma_e=4\times 10^4$. However, this value drops nearly linearly for larger plasmoids (red dashed line in the lower right panel). Since $\gamma_{peak}$ cools linearly with time, this trend infers that the plasmoid size grows linearly with time, i.e., a power law of index -1 in Figure \ref{fig:isolatedphys}, consistent with \citet{Sironi2016}. Note that the plasmoid growth in our simulations consists of both the growth as isolated plasmoids and merging with other plasmoids, indicating that both are linear in time.

\begin{figure}
\centering
\includegraphics[width=0.45\textwidth]{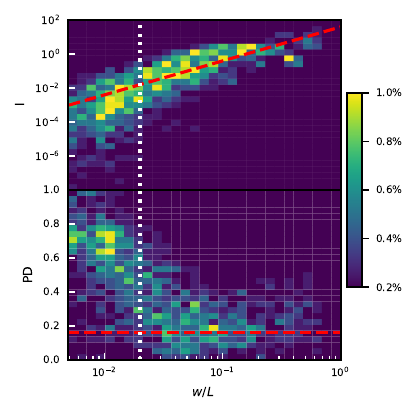}
\caption{2D histograms of radiative properties of isolated plasmoids vs. plasmoid width $w$. From top to bottom, the two panels show the total Stokes $I$ (i.e., luminosity) and the PD. The white dotted line marks $w=0.02 L$ as in Figure~\ref{fig:isolatedphys}. The red dashed lines represent the theoretical values discussed in this section.}
\label{fig:isolatedrad}
\end{figure}

The luminosity of plasmoids generally scales with the square of the plasmoid width, as shown in the red dashed line in Figure \ref{fig:isolatedrad} upper panel. This suggests that the emissivity in plasmoids is roughly constant regardless of the plasmoid size. We note that the luminosity of small plasmoids (smaller than the white dotted line) tends to deviate to slightly lower luminosity than the red dashed line, even though they have the highest $\gamma_{peak}$. This is because the small plasmoids have relatively weaker magnetic fields since they preferentially stay in regions with less dense magnetic field lines as mentioned above. As a result, their luminosity is slightly lower.

Due to the quasi-circular shape of the plasmoids, they yield a nontrivial PD. In our simulations, the viewing angle is along $+y$, the expected PD can be estimated via the following integration in an ellipse. For simplicity, we assume that the magnetic field strengths and nonthermal electron distributions are constant in the quasi-circular plasmoid, then the integration of Stokes $I$, $Q$, $U$ are given by
\begin{equation}
\begin{aligned}
S &= \int^{2\pi}_0 \int^{r(\theta)}_0 dr d\theta\, r s ~~,
\end{aligned}
\end{equation}
where $s=\epsilon$ for Stokes $I$, $s=\epsilon \Pi \cos{2\alpha}$ for Stokes $Q$, and $s=\epsilon \Pi \sin{2\alpha}$ for Stokes $U$, and $\Pi=70\%$ is assumed to be a constant for the synchrotron PD at every point within the ellipse as the electron power-law distribution has an index of $p=-2$, while $\alpha$ is the angle between the line perpendicular to the tangent of the ellipse at $\theta$ and $x$-axis. Since $s$ does not depend on $r$, we have
\begin{equation}
\begin{aligned}
S &= \frac{1}{2} \int^{2\pi}_0 d\theta\, r^2 s ~~,
\end{aligned}
\end{equation}
where $r=ab/\sqrt{(b\cos{\theta})^2+(a\sin{\theta})^2}$ is the radius of the ellipse from the center at different $\theta$ in which $a=w_{\parallel}/2$ and $b=w/2$. Therefore, we have
\begin{equation}
\begin{aligned}
I & = \frac{1}{2} \int^{2\pi}_0 d\theta\, r^2 \epsilon &= & \frac{\pi}{4} \epsilon w_{\parallel}w \\
Q & = \frac{1}{2} \int^{2\pi}_0 d\theta\, r^2 \epsilon \Pi \cos{2\alpha} &=& \frac{1}{2} \epsilon \Pi \int^{2\pi}_0 d\theta\, r^2 \cos{2\alpha} \\
U & = \frac{1}{2} \int^{2\pi}_0 d\theta\, r^2 \epsilon \Pi \sin{2\alpha} &=& \frac{1}{2} \epsilon \Pi \int^{2\pi}_0 d\theta\, r^2 \sin{2\alpha} 
\end{aligned}~~,
\end{equation}
With some math one can find $\alpha$ as
\begin{equation}
\tan{(\alpha+\frac{\pi}{2})}=\frac{b^2}{a^2}\tan{(\theta+\frac{\pi}{2})} ~~.
\end{equation}
Then Stokes $Q$ and $U$ are given by
\begin{equation}
\begin{aligned}
Q &=\frac{1}{2} \epsilon \Pi \int^{2\pi}_0 d\theta\, \frac{a^2b^2}{(b\cos{\theta})^2+(a\sin{\theta})^2} \frac{b^4\cos^2\theta-a^4\sin^2\theta}{b^4\cos^2\theta+a^4\sin^2\theta} \\
U &=\frac{1}{2} \epsilon \Pi \int^{2\pi}_0 d\theta\, \frac{a^2b^2}{(b\cos{\theta})^2+(a\sin{\theta})^2} \frac{2a^2b^2\sin\theta\cos\theta}{b^4\cos^2\theta+a^4\sin^2\theta}
\end{aligned} ~~.
\end{equation}
By observing the expression for Stokes $U$, the integral is proportional to $\tan\theta$, thus the integration is zero. The Stokes $Q$ integration can be done analytically by noticing that $d\tan\theta=\sec^2\theta d\theta$, and we use $z=\tan\theta$, then we have
\begin{equation}
\begin{aligned}
Q &=2\epsilon \Pi \int^{\infty}_0 dz\, \frac{a^2b^2}{b^2+a^2z^2} \frac{b^4-a^4z^2}{b^4+a^4z^2} &=& \epsilon \Pi \frac{\pi ab(b-a)}{a+b}
\end{aligned} ~~.
\end{equation}
Thus we have
\begin{equation}
\begin{aligned}
PD=\frac{\sqrt{Q^2+U^2}}{I}=\frac{|a-b|}{a+b}\Pi ~~.
\end{aligned}
\end{equation}
For $w_{\parallel}=1.6w$, we find that PD is about 16\%. If we view from a different angle, the Stokes $Q$ and $U$ in the above equation need to be corrected by a projection factor. The net PD is zero only if we are viewing in the $xy$ plane along an angle $\sim \arctan{1/1.6}$ from $x$ axis, where the projected ellipse becomes a perfect circle. In practice, the nonthermal particles and magnetic field distributions inside isolated plasmoids can change in time, causing fluctuations in PD and PA as shown in the previous section. The above PD estimate should be considered as the average value. The PD deviates from the theoretical value for small plasmoids as shown in Figure \ref{fig:isolatedrad} lower right panel. This is because our radiation transfer code has a lower resolution than PIC. For plasmoids smaller than $w=0.02L$ (the white vertical line), the plasmoid $w$ and $w_{\parallel}$ are only resolved by $\sim 3$ radiation cells, less than 9 cells for the plasmoid area. This should yield a PD of $\sim 1/\sqrt{9} \times 70\% =23.3\%$ even if the magnetic fields are completely random in the plasmoids. The smallest plasmoids contain only one radiation cell, resulting in the maximal PD $\sim 70\%$.

\section{Merging plasmoids}
\label{sec:merger}

\begin{figure*}
\centering
\includegraphics[width=0.8\textwidth]{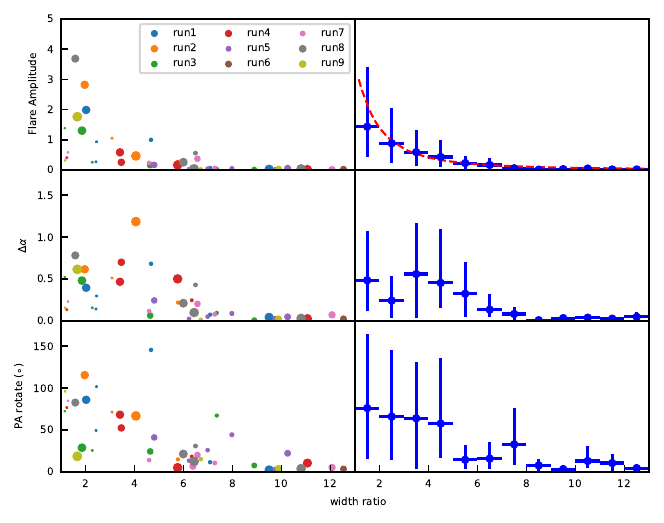}
\caption{Relative flare amplitudes, spectral index changes, and the absolute value of PA variation amplitudes due to binary mergers, plotted against the width ratio of the merging plasmoids. Left panels are the distributions of each quantity. The size of the circles is proportional to the size of the post-merger plasmoids. A few points that are beyond the range of the y-axis are not shown. Different colors correspond to different runs. Right panels are the binned distribution based on the skewed Gaussian distribution.}
\label{fig:merger}
\end{figure*}

\begin{figure}
\centering
\includegraphics[width=0.45\textwidth]{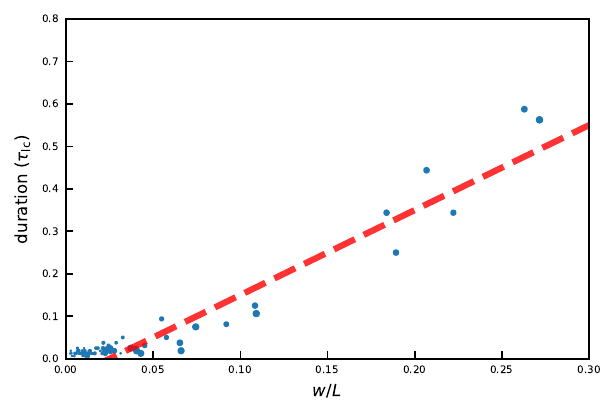}
\caption{Duration of a merger plotted against the characteristic length of the secondary current sheet. The red dashed line corresponds to a fitting that is consistent with a growth rate of $0.1c$.}
\label{fig:growth}
\end{figure}

This section presents the radiative properties of binary plasmoid mergers. We find that the flare amplitude is roughly anti-proportional to the size ratio of the merging plasmoids. Additionally, the optical photon spectral index becomes harder by $\sim 0.5$ during the merger if the width ratio of the merging plasmoids is $R_w\lesssim 5$. Similarly, PA variations only occur with the same width ratio, whose amplitudes can lie between 0 and $180^{\circ}$ with an average of $90^{\circ}$. Finally, the merging time scale is proportional to the size of the secondary current sheet, with a rate of $\sim 0.1 c$.

As shown in the previous section, large plasmoids have more energy than small plasmoids. As a result, mergers between relatively large plasmoids generally dissipate more energy and radiate more strongly than mergers between relatively small plasmoids. Here we choose to study the relative changes in the flux, optical photon spectral index, and PA to avoid differences due to the size of plasmoids participating in the merger. Additionally, we study the dependence of these changes on the width ratio of the merging plasmoids instead of their widths. Figure \ref{fig:merger} (left panels) shows the results\footnote{Several mergers with width ratio $R_w> 13$ are not shown as they lead to no changes in flux or polarization.}. With the above choices, we find that indeed as long as the width ratio is similar, mergers leading to large post-merger plasmoids, which happen between relatively large plasmoids, yield similar variations in radiation signatures compared to those making small post-merger plasmoids (except for $\Delta \alpha$, where the small plasmoid mergers show smaller spectral changes than large ones). Additionally, Runs 2 and 3 with different $\sigma$ and Runs 8 and 9 with different cooling factors do not show clear difference from the other runs, since we only consider the relative changes in the flux and polarization during plasmoid mergers. Therefore, we only consider the width ratio parameter here and include all runs to boost statistics in the number of mergers. To better illustrate the statistical results, we bin the mergers based on the width ratio, and the right panels of Figure \ref{fig:merger} show a skewed Gaussian distribution for the statistics: the 0.84, 0.5, and 0.16 quantiles of the skewed Gaussian distribution are plotted as the upper y error bar, median, and lower y error bar, respectively.

We find that only mergers with a width ratio $R_w\lesssim 3$ can produce a flare with an amplitude greater than 1 (i.e., double the flux). Mergers with a large width ratio $R_w\gtrsim8$ bring about trivial flare regardless of the size of participating plasmoids. Interestingly, the flare amplitude seems to follow an anti-proportional relation to the width ratio. This can be understood in the following way. As shown in the previous section, isolated plasmoids have approximately the same emissivity. For a binary merger, say the larger participating 2D plasmoid has a surface area of $A$, then that of the smaller plasmoid is $\frac{1}{R_w^2} A$, where $R_w$ is the width ratio of the larger plasmoid over the smaller one. Therefore, the luminosity of the two isolated plasmoids right before the merger starts is given by
\begin{equation}
L_0\propto (1+\frac{1}{R_w^2}) A~~.
\end{equation}
When the two plasmoids start to merge, the length of the contact region obviously cannot be larger than the width of the smaller plasmoid, thus the size of the secondary reconnection layer is at most proportional to the size of the smaller plasmoid. Since the magnetic energy density is approximately the same for plasmoids of any sizes, if a constant portion of the magnetic energy in the secondary reconnection layer is dissipated and transformed to radiation, then the extra luminosity induced by the merger should be $L_1=C \frac{1}{R_w^2} A$, where $C$ describes a portion of the magnetic energy of the smaller plasmoid that is dissipated. As a result, the flare amplitude is given by
\begin{equation}
\frac{L_1}{L_0}\propto \frac{(1/R_w^2) A}{(1+1/R_w^2)A} = \frac{1}{(1+R_w^2)}~~.
\end{equation}
This fitting curve is shown in red in the upper right panel of Figure \ref{fig:merger}. It is clear that this model describes the flare amplitude very well. However, it deviates slightly from the statistics for mergers of width ratio $R_w \sim 1$. This is because for $R_w\sim 1$, the secondary reconnection region is no longer proportional to the size of the smaller plasmoid, but it has to be smaller than that. As a result, $L_1<C \frac{1}{R_w^2} A$, so that the actual flare level is lower than the theoretical value of the red curve in Figure~\ref{fig:merger}.

The optical photon spectral index hardens by $\sim 0.5$ for $R_w\lesssim 5$. Along with the flare amplitude, plasmoid mergers with $R_w\lesssim 3$ can well explain the often observed harder-when-brighter trend. This hardening feature quickly disappears for $R_w>5$. Obviously, if the flare level is too low, generally we do not expect to see spectral hardening. But to understand this behavior in more detail, we compare the electron spectrum at the beginning of the merger and at the flare peak of the merger. The main difference is that at the flare peak, the electron spectrum nearly extends to the electron magnetization factor $\sigma_e$ with an index of $p\sim 2$, as shown in the upper panel of Figure \ref{fig:spectra_run1_idx45} (blue is near the beginning while the red curve is near the peak). This suggests that regardless of the width ratio of the participating plasmoids, mergers result in a hard spectrum of newly accelerated particles. However, if the width ratio is too large, the total number of newly accelerated particles is smaller than the number of particles already existing in the plasmoids participating in the merger, then the total particle spectral shape is trivially affected by the newly accelerated particles during the merger. This explains why $\Delta \alpha$ quickly drops to zero for width ratio $R_w>5$. For $R_w\lesssim 5$, $\Delta \alpha$ has a relatively large range because of the spectrum in the plasmoids. As shown in the previous section, small plasmoids, which are newly generated, have an electron spectrum extend to $\gamma_e\sim 10^4$, similar to the particle spectrum accelerated during mergers. However, large plasmoids, due to cooling, have lower spectral cutoffs. Since the optical synchrotron emission is dominated by electrons at $\gamma_e\sim 10^4$ in our setup, large plasmoids have the optical emission deep in the radiatively cooled spectrum. Consequently, when they merge, the newly accelerated particles, which are not yet cooled, push the optical photon spectrum much harder. Small plasmoids, on the other hand, have hard optical photon spectra before the merger anyway, thus the merger does not harden the spectral index significantly. This explains why mergers between small plasmoids tend to occupy a region with smaller $\Delta \alpha$ even for $R_w\lesssim 5$ in Figure~\ref{fig:merger} middle left panel.

PA variations are only considerable for $R_w\lesssim 5$ for similar reasons: PA changes are induced by newly accelerated particles; if they are not enough to affect the nontrivial PD resulting from the quasi-circular shape of the plasmoids, PA will not change significantly. Most of the PA variations are $\sim 90^{\circ}$. The reason is that the plasmoid mergers typically start from head-on collision between two plasmoids, so that the newly accelerated particles are at the secondary reconnection layer that is perpendicular to the primary current sheet. The magnetic field lines in the secondary current sheet are then parallel to the net polarization direction of the isolated plasmoids, thus the newly accelerated particles preferentially emit synchrotron polarized in the $\sim 90^{\circ}$ from the polarization direction of the pre-merger plasmoids. To complete a full $\sim 180^{\circ}$ PA rotation, the merger needs to meet three criteria: the merger cannot complete before the two plasmoids rotate $180^{\circ}$ from each other; the newly accelerated particles at the secondary current sheet cannot be mostly cooled before the rotation completes; the secondary current sheet cannot generate too disordered magnetic fields. For instance, the merger shown in Section \ref{sec:method} fails to make the $\sim 180^{\circ}$ PA rotation because of the very disordered magnetic field in the secondary reconnection region. Therefore, flares driven by reconnection are not necessarily accompanied by large PA swings.

Figure \ref{fig:growth} shows the growth rate due to the merger. This is simply the width of the post-merger plasmoid minus the width of the larger plasmoid that participates in the merger, plotted against the duration of the merger defined in Section \ref{sec:method}. Due to the limited number of mergers between large plasmoids, we do not have very good statistics for the growth rate. But Figure \ref{fig:growth} can be fit with a linear growth rate that is $\sim 0.1 c$, consistent with the growth rate of isolated plasmoids found in \citet{Sironi2016}.

\section{Summary and Discussion}
\label{sec:summary}

To summarize, we statistically study the radiative properties of isolated and merging plasmoids in magnetic reconnection. Our results are based on combined 2D PIC and polarized radiation transfer simulations. Our image processing method that combines the magnetic vector potential and particle density can clearly identify the boundaries of isolated and merging plasmoids. Our key findings are as follows.
\begin{enumerate}
\item Radiation from magnetic reconnection is dominated by isolated and merging plasmoids. The contribution from the primary current sheet is mostly negligible.
\item The luminosity of isolated plasmoids is proportional to its size.
\item Isolated plasmoids have comparable magnetic energy density and particle kinetic energy, but the former is slightly higher.
\item Isolated plasmoids generally have non-trivial polarization due to their quasi-circular morphology.
\item Flares driven by plasmoid mergers have amplitudes that are anti-proportional to the size ratio of the plasmoids participating in the mergers.
\item Mergers between plasmoids with a relatively small width ratio ($R_w\lesssim 5$) can lead to spectral hardening near the synchrotron spectral peak.
\item Mergers between plasmoids with relatively small width ratio ($R_w\lesssim 5$) can produce PA variations ranging from 0 to $\sim 180^{\circ}$, with an average of $\sim 90^{\circ}$.
\item PA swings are due to nonthermal particles that stream along the magnetic field lines surrounding the merging plasmoids, which result from the merging plasmoids rotating around each other.
\item The plasmoid growth rate, regardless of growing in an isolated environment or through merging, is $\sim 0.1 c$.
\end{enumerate}

Our method clearly shows that plasmoids and mergers can be identified by multiple physical quantities (in our paper they are the magnetic vector potential and particle density). Although our segmentation is done manually, the identified plasmoids and mergers and their evolution can be fed into machine learning to segment plasmoids and mergers automatically. This can significantly reduce the computational cost and human labor involved in our present study, because right now we have to manually track the plasmoid evolution and merging processes case by case and code the envelops accordingly. In the future with machine learning, we can perform the above studies on more and longer combined PIC and radiation transfer simulations to boost the statistics of plasmoids and mergers, thus reducing the systematic uncertainties in modeling the radiation and polarization signatures from magnetic reconnection.

Our results are readily applicable to model 2D reconnection. However, three more steps are necessary to better study the multi-wavelength radiation signatures from magnetic reconnection. Firstly, we only study the synchrotron emission in this paper, using the optical band that is close to the cooling break as an example. \citet{ZHC2020} shows that synchrotron emission at different parts of the spectrum can behave differently in both the flux and polarization. Additionally, \citet{ZHC2022} finds that the synchrotron self Compton and external Compton exhibit different light curves compared to the synchrotron emission. In particular, the synchrotron self Compton can produce very fast flares due to the highly inhomogeneous particle and photon distributions in the secondary reconnection layer formed by plasmoid mergers. Although the present work can study all the above signatures, we do not have adequate statistics to reach any conclusive results. Therefore, we plan to investigate the multi-wavelength patterns with future machine learning studies. Secondly, our results are limited to 2D reconnection. 3D PIC simulations have shown that turbulence can become important in magnetic reconnection \citep{Guo2021} and that substantial particle acceleration takes place very efficiently in the reconnection upstream \citep{Giannios2010,ZSG2021,ZSGP2023}. Although 2D slices along the guide field direction of the 3D simulations appear similar to 2D reconnection, the flux ropes, which are the 3D counterparts of plasmoids, can have much more complicated structures. Consequently, their mergers are much more complicated as well. Moreover, 3D simulations naturally allow different viewing angles, which can have a major impact on radiation and polarization. It is therefore necessary to examine that the radiation signatures obtained from 2D simulations are statistically consistent with those from the 3D simulations. Finally, PIC simulations cannot extend to realistic blazar zone scales with the available computing power, thus the initial and boundary conditions of the reconnection region may be very different from the realistic situation. Radiation signatures based on PIC simulations have to be tested against future hybrid simulations involving fluid dynamics, particle transport, and radiation transfer.

\bibliographystyle{mnras} 
\bibliography{merger.bib} 

\appendix
\section*{Data Availability}
The data underlying this article will be shared on reasonable request to the authors.

\section*{Acknowledgements} 

We thank the referee for very helpful comments. HZ is supported by NASA under award number 80GSFC21M0002. HZ's work is supported by Fermi GI program cycle 16 under the award number 22-FERMI22-0015. LD and DG acknowledge support from the NSF AST-2107802, AST-2107806 and AST-2308090 grants.

\label{lastpage}
\end{document}